\documentclass[twocolumn]{emulateapj}

\usepackage{graphicx}
\usepackage{amssymb}
\usepackage{epsfig}
\usepackage{amsmath}
\usepackage{bm}

\newcommand\be{\begin{eqnarray}}
\newcommand\ee{\end{eqnarray}}
\newcommand\lsim{\mathrel{\rlap{\lower3pt\hbox{\hskip1pt$\sim$}}
     \raise1pt\hbox{$<$}}} %less than or approx. symbol
\newcommand\gsim{\mathrel{\rlap{\lower3pt\hbox{\hskip1pt$\sim$}}
     \raise1pt\hbox{$>$}}} %greater than or approx. symbol

\shorttitle{Lifetime of GRBs}
\shortauthors{Jeong \& Lee}

\begin{document}

\title[]{Red-Shift Distribution of Gamma-ray Bursts and Their Progenitors}

\author{Soomin Jeong}
\affil{
Research Center of MEMS Space Telescope,
Dept. of Physics, Ewha Womans University, Seoul 120-750, Korea \\
Department of Physics, Pusan National University, Busan 609-735, Korea}
\email{smjeong@hess.ewha.ac.kr}
\and
\author{Chang-Hwan Lee}
\affil{Department of Physics, Pusan National University, Busan 609-735, Korea}
\email{clee@pusan.ac.kr}
%\date[]{Received July 20  2007}

\begin{abstract}
Gamma ray bursts have been divided into two classes, long-soft gamma ray burst and short-hard gamma ray burst according to the bimodal distribution in duration time. Due to the harder spectrum and the lack of afterglows of short-hard bursts in optical and radio observations, different progenitors for short-hard bursts and long-soft bursts have been suggested. Based on the X-ray afterglow observation and the cumulative red-shift distribution of short-hard bursts, \citet{Nak06} found that the progenitors of short-hard bursts are consistent with old populations, such as mergers of binary neutron stars.
Recently, the existence of two subclasses in long-soft bursts has been suggested after considering multiple characteristics of gamma-ray bursts, including fluences and the duration time.

In this work, we extended the analysis of cumulative red-shift distribution to two possible subclasses in L-GRBs. We found that two possible subclass GRBs show different red-shift distributions, especially for red-shifts $z > 1$. Our results indicate that the accumulative red-shift distribution can be used as a tool to constrain the progenitor characteristics of possible subclasses in L-GRBs.
\end{abstract}

%\pacs{84.40.Ik, 84.40.Fe}

\keywords{gamma rays: bursts --- binaries: close --- supernovae: general}

\section{INTRODUCTION}

Traditionally, gamma-ray bursts (denoted as GRBs) have been divided into two classes which are separated by the duration time of gamma-ray bursts $T_{90}=2$ sec \citep{Kou93}. Since the short-duration gamma-ray bursts had relatively hard observed spectra than long-duration gamma-ray bursts, two classes are usually called as long-soft gamma ray burst (denoted as L-GRB) and short-hard gamma-ray burst (denoted as SHB).
From the afterglow observations of L-GRBs, the association between the L-GRBs and supernovae/hypernovae has been suggested. These observations suggest that the progenitors of L-GRBs are massive giant stars which are collapsing.
%One of the best can \citep{Wos93,Mac99}. Lee
One of the best candidate is the Collapsar model \citep{Woo93,Mac99} in which rapidly rotating black hole with Kerr parameter about 0.8 is formed after GRB. Recently, spinning black holes with high Kerr parameters, $a_\star \approx 0.8$, have been observed in soft X-ray black hole binaries \citep{McC06,Sha06}. These observations support the idea that the rapidly rotating black holes in soft X-ray black hole binaries are the remnants of L-GRBs \citep{Bro00,Lee02}, providing the natural sources for Collapsars.
%We believe that the massive collapsing giants which leave black holes behind are the progenitors of %some L-GRBs.
%

Due to the differences in the observational characteristics of SHBs compared to those of L-GRBs, such as lack of afterglow observations in optical and radio band and harder spectrum, etc., different progenitors for SHBs have been suggested. Mergers of binary neutron stars \citep{Eic89,Nar92} are among the best candidates. After recent X-ray afterglow observations by Swift and HETE-II, the associations between SHBs and the host galaxies were reported (see Table~\ref{tab1} and references there in).
Based on the X-ray afterglow observation of SHBs and the cumulative red-shift distribution of SHBs, \citet{Nak06} found that the progenitors of SHBs are consistent with old populations. This finding supports mergers of binary neutron stars as progenitors of SHBs.

Recently, the existence of two subclasses in L-GRBs has been suggested after considering both fluences and the duration time \citep{Cha07}. They divided L-GRBs into Clusters II and III which were separated by the fluences combined with the duration time. They classified L-GRBs with higher fluences as Cluster III and suggested that their progenitors are the massive collapsing stars, such as Collapsar \citep{Woo93,Mac99}. As a possible origin of Cluster II GRBs, they suggested neutron-star, white-dwarf binaries. If there exist two subclasses in L-GRBs as \citet{Cha07} suggested and only Cluster III GRBs are associated with massive collapsing giants which are the progenitors of supernovae or hypernovae, the observed red-shift distribution of two subclasses might show different behaviour.

In this work, we extended the work of \citet{Nak06} to two possible subclasses in L-GRBs.
We found that Clustters II and III show different red-shift distribution at high red-shift $z>1$.
Our results indicate that the accumulative red-shift distribution can be used as a tool to constrain the lifetimes of possible subclasses in GRBs.
In Sec.~\ref{sec-rate}, the numerical methods which we used in our analysis are summarized \citep{Nak06}. The estimated lifetime of SHBs are summarized in Sec.~\ref{sec-SHB}.
We confirmed that the progenitors of SHBs are consistent with old population with lifetime $\tau_\ast=6.5$ Gyr, indicating that SHBs are from the mergers of binary neutron stars.
In Sec.~\ref{sec-LGRB}, the cumulative red-shift distributions of two possible subclasses in L-GRB have been summarized. Our results show that there are clear differences in the red-shift distributions of two possible subclasses in L-GRBs for the red-shift  $z>1$. This suggests that the cumulative red-shift distribution can be used as a tool to distinguish the subclasses in L-GRBs.
Our final conclusion follows in Sec.~\ref{sec-con}.

\section{Gamma-Ray Bursts Observation Rate}
\label{sec-rate}

\begin{figure}[b]
\begin{center}
\includegraphics[width=8.5cm]{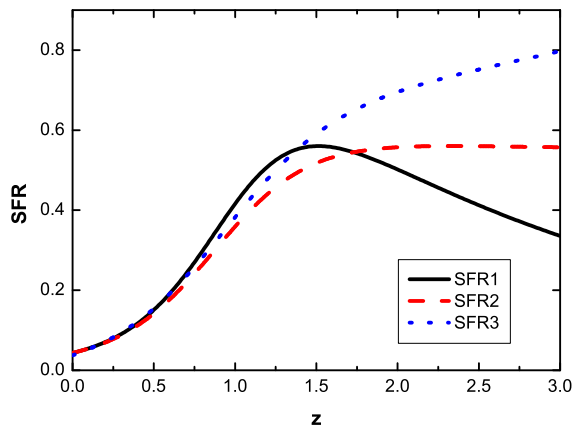}
\end{center}
\caption{Normalized star formation rate (SFR) as a function of red-shift $z$. Due to the uncertainty in the estimation of SFR at high red-shift, $z>1.5$, three different choices of parametrization, Eq.~(\ref{eq-SFR}),  are used following \citet{Por01}. }
\label{fig1}
\end{figure}

To examine the progenitor's life time and the local rate of GRBs, we used the local  GRB rate suggested by \citet{Nak06},
%
%\begin{widetext}
\begin{equation}
{R}_{\rm GRB}(z)\propto\int_{z}^{\infty}{\rm SFR}(z\prime)f(t(z)-t(z\prime))\frac{dt}{dz\prime}
dz\prime
\label{eq-RSHB}
\end{equation}
%\end{widetext}
%
where ${R}_{\rm GRB}$(z) is the intrinsic GRB rate (per unit comoving volume and comoving time),
${\rm SFR}(z)$ is the local star formation rate (SFR) at red-shift z (per unit comoving volume and comoving time), $t(z)$ is the age of universe at red-shift $z$, and $f(\tau)$ is the fraction of GRB progenitors that are born with a lifetime $\tau$.
Note that the absolute value of ${R}_{\rm GRB}$ is not important because we normalized the final cumulative distribution to unity.
In this work, we used three different parameterizations of the global star formation rate
per unit comoving volume following \citet{Por01} and \citet{Nak06}.
\be
{\rm SFR}_1(z) &\propto& \frac{\exp(3.4 z)}{\exp(3.8 z) + 45} \times \frac{H(z)}{(1+z)^{3/2}}
 \nonumber \\
{\rm SFR}_2(z) &\propto& \frac{\exp(3.4 z)}{\exp(3.4 z) + 22} \times \frac{H(z)}{(1+z)^{3/2}}
 \nonumber\\
{\rm SFR}_3(z) &\propto& \frac{\exp(3.05 z - 0.4)}{\exp(2.93 z) + 15} \times \frac{H(z)}{(1+z)^{3/2}}
\label{eq-SFR}
\ee
where $H(z)$ in the standard $\Lambda$CDM Universe model \citep{Cop06} is
\be
H(z) = H_0 \left[\Omega_m (1+z)^3 + \Omega_\Lambda\right]^{1/2}
\label{eq-H}
\ee
with Hubble constant $H_0$. In Fig.~\ref{fig1}, we summarized the normalized SFR.
One can see the clear differences of SFR at high red-shift $z>1.5$ \citep{Por01}.

We used two different forms of $f(\tau)$ in Eq.~(\ref{eq-RSHB}) following \citet{Nak06},
\be
f_1(\tau) &\propto & \tau^{-\eta}
\label{eq-power} \\
f_2(\tau) &\propto &
\frac{1}{\tau\sigma\sqrt{2\pi}}\exp\left(-\frac{\left[\ln(\tau)-\ln(\tau_{\ast})\right]^{2}}{2\sigma^{2}}\right).
\label{eq-log}
\ee
The first is the power-law distribution with which the dominant portion of GRBs were produced at the early evolutionary stage of their progenitors.
The latter is the log-normal distribution in which $\tau_{\ast}$ is the mean time and $\sigma$ is the dispersion. In this work we used $\sigma=0.3$ \citep{Nak06}.

In order to compare the local GRB birth rate at red-shift $z$ with the current observation on earth, we used the differential GRB observation rate $d \dot{N}_{obs}/dz$;
\begin{equation}
\frac{d\dot{N}_{obs}}{dz} = \left[4\pi d_{L}^{2}k(z)\right]^{1-\beta}\phi_{0}\frac{{R}_{\rm GRB}}{1+z}\frac{dV}{dz}\int P^{-\beta}S(P)dP
\label{eq-dNdz}
\end{equation}
where the luminosity distance $d_L(z)$ is
\be
d_L(z) = (1+z) \int_0^z \frac{1}{H(z^\prime)} dz^\prime
\ee
and the $k(z)$ correction, which includes the conversion from energy flux to photon flux, is  \citep{Nak06}
\be
k(z)  \approx 2\times 10^{-7} (1+z)^{-1.5}  \ {\rm ergs}.
\ee
$S(P)$ is the probability for detecting the bursts with a peak photon flux $P$ which in turn depends on the luminosity $L$ and the red-shift $z$ as well as on the spectrum bursts. Since the exact form of $S(P)$ is unknown, in this work we used the simplest form $S(P)=1$. The parameter $\beta$ in Eq.~(\ref{eq-dNdz}) is the power in the luminosity function $\phi(L)=\phi_{0}L^{-\beta}$ and we used $\beta=2$ which is the best fit parameter found by \citet{Nak06}.

\section{Lifetime of Short-Hard Bursts}
\label{sec-SHB}

The observed red-shifts of SHBs are summarized in Table~\ref{tab1}. For GRB 050813, two values of red-shifts are given due to the uncertainty in the association between the putative host galaxy and the SHB \citep{Pro05,Ber06}. In our analysis, we used 7 SHBs without GRB060502B in which the association with the putative host galaxy is uncertain. Our final conclusion on the lifetime of SHBs remains the same even if we include GRB060502B.

\begin{table}
\begin{center}
%\begin{ruledtabular}
%\centerline{ Red-shifts of SHBs}
\begin{tabular}[c]{ccccc}
\hline {SHB}&   Red-shift (\textbf{z})  & Type & Association$^\dagger$ & Ref. \\
\hline
050509B & 0.225 & E(c) & $3-4\sigma$ & 1 \\
050709 & 0.16 & Sb/c & Secure & 2 \\
050724 & 0.258 & E/S0 & Secure & 3 \\
050813 & 0.72 or 1.8  & E/S0 &  & 4\\
051221A & 0.546 & Late & Secure & 5\\
060502B(?) & 0.287(?) & Early & $\approx 90\%$ & 6 \\
790613 & 0.09 & E/S0 & $3\sigma$ & 7 \\
000607 & 0.14 & Sb  & $2\sigma$ & 7 \\
\hline
\end{tabular}
%\end{ruledtabular}
\caption{Putative host galaxy properties of SHBs \citep{Nak07}. References: 1) \citet{Geh05,Fox05,Blo06}; 2) \citet{Fox05,Pro05,Cov06}; 3) \citet{Ber05}; 4) \citet{Pro05,Ber06};
5) \citet{Sod06}; 6) \citet{Blo07}; 7) \citet{Gal05}. $\dagger$ The confidence level of the association between the putative host and the SHB.} \label{tab1}
\end{center}
\end{table}

\begin{figure}
\begin{center}
\includegraphics[width=8.5cm]{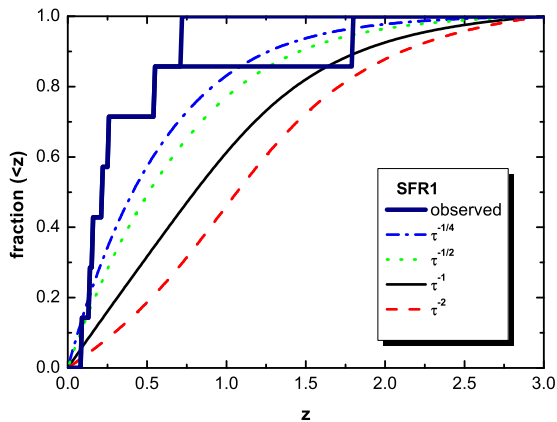}

\includegraphics[width=8.5cm]{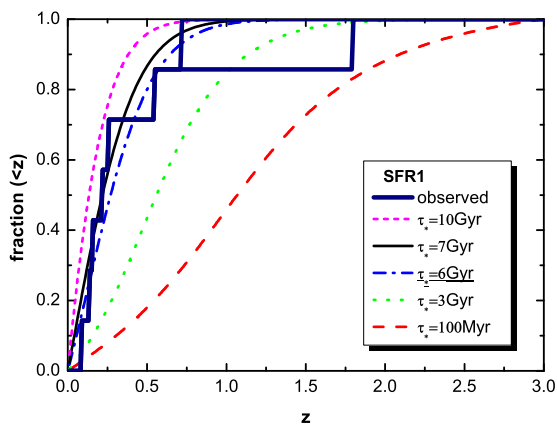}
\end{center}
\caption{The cumulative red-shift distribution of SHBs for two different choices of $f(\tau)$. Here we used SFR1 and $\Omega_m=0.3$ for the matter contents of the Universe.
The upper panel corresponds to power-law distribution $f(\tau)=\tau^{-\eta}$ and the lower panel corresponds to the log-normal distribution, Eq.~(\ref{eq-log}).
%Note that SFR1 and luminosity function $\phi(L) \propto L^{-2}$ are used in all calculation.
Thick solid line corresponds to the observed data. Two different curves are the consequence of the uncertainty in the red-shift estimation of SHB050813.}
\label{fig2}
\end{figure}

The cumulative red-shift distribution of SHBs are summarized in Fig.~\ref{fig2}.
From the upper panel in this figure, one can see that it's very hard to fit the observed distribution using power-law distribution. We also confirmed that SFR2 and SFR3 produce similar results.
From these results, one can confirm that SHBs are not associated with explosive phenomena of giants, such as supernovae or hypernovae. In the lower panel of Fig.~\ref{fig2}, the lognormal distribution Eq.~(\ref{eq-log}) with various mean time $\tau_\ast$ is considered. We found that the old population with $\tau_\ast \approx 6.5$ Gyr gives the best fit to the observation.
This results indicates that SHBs are associated with the mergers of neutron star or black hole binaries \citep{Nak07,Bet07,Lee07}.

%\section{Dependence on the cosmology models}
%\label{sec-cos}

\begin{figure}[t]
\begin{center}
\includegraphics[width=8.5cm]{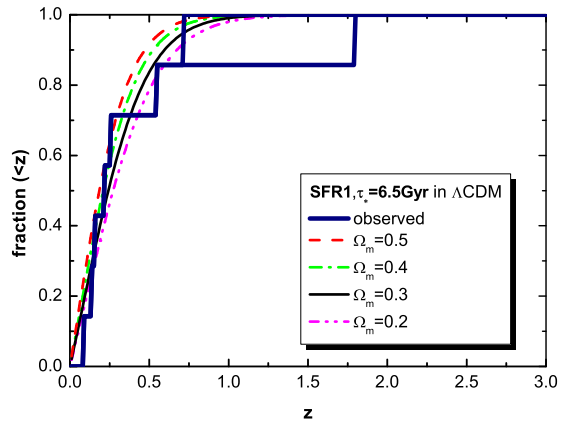}

\includegraphics[width=8.5cm]{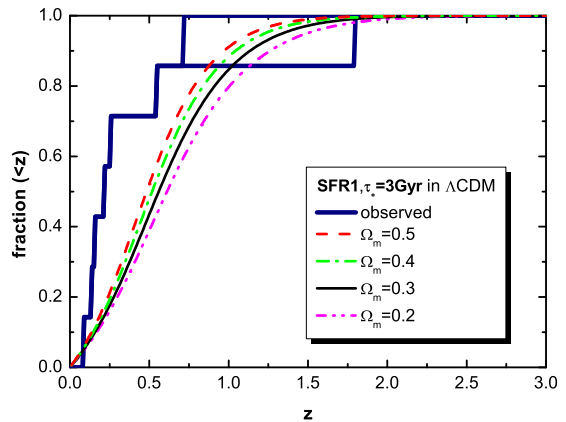}
\end{center}
\caption{The cumulative red-shift distribution of SHBs with SFR1 for different choices of $\Omega_m$ in $\Lambda$CDM model. Two different mean values of delay time, 6.5 Gyr and 3 Gyr, are used for the comparison. } \label{fig3}
\end{figure}

In previous results, we used the matter content in the Universe $\Omega_m=0.3$ which is close to the current observation.
We investigated SHB progenitor life time by changing the matter contents $\Omega_m$ in Eq.~(\ref{eq-H}). The results are summarized in Fig.~\ref{fig3} in which we considered two different choices of the mean time $\tau_\ast$. From Fig.~\ref{fig3}, one can see that the old population with mean time $\tau_\ast = 6.5$ Gyr is consistent with the observation independently of the matter contents. We also tested other Universe models, such as Phantom CDM model, and found that our conclusion is independent of the particular Universe models. This observation is consistent with the fact that most of the observed SHBs are distributed in low red-shifts.

%------------------------------------------------
%------------------------------------------------
%------------------------------------------------

\section{Subclasses in Long-Soft Gamma-Ray Bursts}
\label{sec-LGRB}

Due to the earlier afterglow observations, the progenitors of L-GRBs are believed to be related with those of supernovae or hypernovae. However, recently the possibility of two sub-classes in L-GRBs was suggested by \citet{Cha07}.
They divided long-duration GRBs into Cluster II for low fluence and Cluster III for high fluence, which were separated by the line,
\be
F_T = 1.6 \times 10^{-4}/T_{90}\ {\rm ergs/cm^2}
\label{eq-ft}
\ee
where $F_T$ is the fluence and $T_{90}$ is the time corresponding to 90\% of the flux.
They showed that Cluster II GRBs have nearly constant isotropic energy output of $10^{52}$ ergs while Cluster III GRBs have isotropic energy in wide ranges from $10^{52}$ to $10^{54}$ ergs. As possible origins, they suggested mergers between neutron star and white dwarf for Cluster II and collapse of massive stars for Cluster III.

\begin{figure}
\begin{center}
\includegraphics[width=8.5cm]{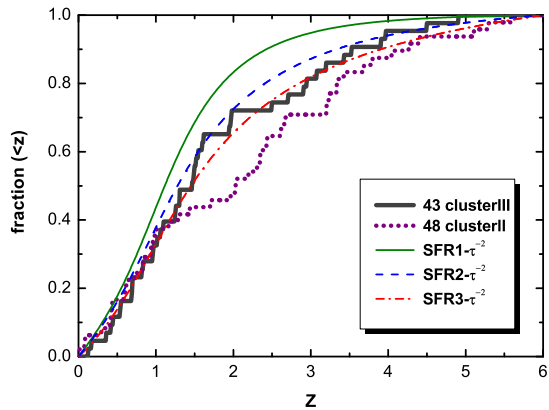}
\end{center}
\caption{The cumulative red-shift distribution of L-GRBs.
Thick lines correspond to the observed distributions. The thick solid line corresponds to the 43 Cluster III GRBs and the thick dotted line corresponds to the 48 Cluster II GRBs. Two clusters show clear difference in the red-shift region $z > 1$.
Thin lines correspond to the lowest cumulative distribution for given star formation rates with $\beta(\tau) =\tau^{-2}$.
} \label{fig6}
\end{figure}

In this work, we investigated the cumulative red-shift distribution of Cluster III using \citet{Nak06} method. If Cluster III is related with the collapse of massive stars, their cumulative distribution should be able to be described by Eq.~(\ref{eq-power}).
The results are summarized in Fig.~\ref{fig6}. In this figure, we tested two different samples of L-GRBs with known red-shifts, 48 Cluster II (thick dotted line) and 43 Cluster III (thick solid line) which were divided by Eq.~(\ref{eq-ft}). Three thin lines are the lowest cumulative distributions for given star formation rates. In Fig.~\ref{fig6}, one can see that independently of SFR models the Cluster II L-GRBs cannot be described by the power-law distribution for the red-shift $z>1$. On the other hand, even though it's too early to make any firm conclusion, distributions with only Cluster III can be easily described by the power-law distribution. This might indicate that there exist two subclasses in the long-duration GRBs.
Main deviations between two clusters in Fig.~\ref{fig6} are in the red-shift region $ z > 1$.
%However, in order to make any firm conclusion, one needs more investigation.

%----------------------------------------------------------------------------

\section{CONCLUSIONS}
\label{sec-con}

In this work, by extending the work of Nakar et al. \citep{Nak06,Nak07}, we investigated the constraints on the progenitor lifetime of SHBs. We confirmed that the SHBs are consistent with old population with mean time $\tau_\ast=6.5$ Gyr. We also confirmed that this conclusion is independent of the details of the cosmology models. This results support the conclusion that the progenitors of SHBs are consistent with old population, such as merging compact star binaries (neutron star--neutron star binaries or neutron star--black hole binaries) \citep{Nak06,Nak07,Bet07,Lee07}.

We also investigated two subclasses Cluster II and III L-GRBs \citep{Cha07} using the same method of \citet{Nak06}. We found that the power-law distribution is more consistent with Cluster III GRBs than with Cluster II GRBs. We also found that the cumulative red-shift distribution of two clusters show clear difference in the red-shift region $z>1$. This result supports the existence of two subclasses in L-GRBs. However, in order to draw firm conclusions, many effects which are not included in our analysis, e.g., the connection between GRBs and the metallicity of the host galaxy, different characteristics of cosmological GRBs and subluminous GRBs, observability premium for different type of GRB progenitors, etc., have to be considered.

\begin{acknowledgments}
We would like to thank Gerald Brown and Ehud Nakar for the helpful discussions.
This work was supported by Creative Research Initiatives (MEMS Space Telescope) of MOST/KOSEF.
\end{acknowledgments}

%\begin{references}

%\end{references}

\end{document}